\documentstyle[11pt]{article}

\def\be{\begin{equation}}
\def\ee{\end{equation}}
\def\bea{\begin{eqnarray}}
\def\eea{\end{eqnarray}}

\newcommand{\ket}[1]{|\kern.3ex#1\kern.3ex\rangle}
\newcommand{\bra}[1]{\langle\kern.3ex #1 \kern.3ex|}
\newcommand{\APPROX}[1]{                
   {{\raisebox{-.3cm}{$\textstyle\simeq$}} \atop {\scriptstyle{#1}}}}
\newcommand{\EXP}[1]{{\mbox{\large e}}^{#1}}
\newcommand{\mean}[1]{\langle#1 \rangle}            
\newcommand{\cotg}{\mathop{\mathrm{cotg}}\nolimits} 

\newcommand{\drond}[2]{\frac{\partial #1}{\partial #2}} 

\newcommand{\sigg}{\sigma_{\mbox{\tiny g}}}
\newcommand{\typ}[1]{#1_{\mbox{\scriptsize typ}}}

\newcommand{\PRL}[3]{#2 {\it Phys. Rev. Lett.} {\bf #1} #3}
\newcommand{\JPhysCM}[3]{#2 {\it J. Phys. Condens. Matter} {\bf #1} #3}
\newcommand{\JdePhys}[3]{#2 {\it J. Physique} {\bf #1} #3} 
 
\newcommand{\PhysRev}[3]{#2 {\it Phys. Rev.} {\bf #1} #3} 
\newcommand{\PhysRep}[3]{#2 {\it Phys. Rep.} {\bf #1} #3} 
\newcommand{\AnnPhys}[3]{#2 {\it Ann. Phys. (N.Y.)} {\bf #1} #3} 
\newcommand{\JPhysA}[3]{#2 {\it J. Phys. A: Math. Gen. } {\bf #1} #3}
\newcommand{\JStat}[3]{#2 {\it J. Stat. Phys  } {\bf #1} #3}
\newcommand{\JPhysI}[3]{#2 {\it J. Phys. I (France)} {\bf #1} #3}

\begin{document}


\begin{center}
{\LARGE On the distribution of the Wigner time delay \\
in one-dimensional disordered systems}

\vspace{1cm}

{\large 
Alain Comtet{\normalsize $\dagger\ddagger$\footnote{
                             E-mail: comtet@ipno.in2p3.fr}}
and Christophe Texier{\normalsize $\dagger$\footnote{
                             E-mail: texier@ipno.in2p3.fr}}}

\vspace{1cm}

$\dagger$ Division de Physique Th\'eorique\footnote{
Unit\'e de recherche des Universit\'es Paris 11 et Paris 6 associ\'ee au CNRS.
}, IPN B\^at. 100, 91406 Orsay C\'edex, France.

$\ddagger$ L.P.T.P.E, Universit\'e Paris 6, 4 place Jussieu, 75252 Paris
C\'edex 05, France.

\end{center}

\begin{abstract}
We consider the scattering by a one-dimensional random potential and derive
the probability distribution of the corresponding Wigner time delay. It is
shown that the limiting distribution is the same for two different models
and coincides with the one predicted by random matrix theory. It is also
shown that the corresponding stochastic process is given by an exponential
functional of the potential.
\end{abstract}

\vspace{2cm}

IPNO/TH 97-15

\newpage


\section{Introduction}

The concept of time delay introduced long ago by Eisenbud and Wigner has
recently received renewed attention. 
In a scattering process the time delay $\tau(k)$ is related to
the time spent in the interaction region by a wavepacket\footnote{
For a detailed and critical analysis of this concept see \cite{buttiker} and
\cite{LandMart}.} with
energy peaked at $E=k^2$. It can be expressed in terms of the derivative of the
$S$ matrix with respect to the energy. If the system is described by a
finite number $N$ of channels, the $N$ time delays are the eigenvalues of
the matrix $-iS^\dagger\drond{S}{E}$. In the context of chaotic scattering a
statistical approach based on random matrix theory allows a determination of the
complete distribution of time delays (for a review see \cite{fyodorov2}).
This problem was first studied in \cite{fyodorov} by a supersymmetric
approach, and in \cite{GopMelBut} by using a statistical analysis.
This latter work provided a physical derivation of the one-channel case for
$\beta=2$ and gave the distribution for other universality classes. It then
recently served as a starting point for \cite{beenakker} where the
$N$-channel distribution is shown to be given by the Laguerre ensemble of
random matrix theory.
A slightly different approach has been developed in 
the context of quasi one-dimensional mesoscopic systems. 
Quite recently a resonant transmission model was
proposed to account for the different behaviours observed in the
metallic and insulating regimes \cite{jalabert}.

In this paper we consider a one-dimensional model with a random potential
whose support is a finite segment of length $L$. A hard wall condition at
the origin reduces the problem to a scattering problem on the half line.
Using standard methods we write down a set of two coupled stochastic
differential equations satisfied by the phase and its derivative with
respect to the energy which is related to the time delay. In the high energy
(or weak disorder) limit we may take advantage of the fact that the phase is
a rapid variable with a uniform distribution. One finally ends up with a
stochastic differential equation for the time delay which coincides
with the one derived in \cite{kumar,heinrichs} and which can be integrated
explicitly in the case where the potential is  white noise.
The resulting expression is given in terms of an exponential
functional of a Brownian motion. An analysis of the same model with a
different type of disorder yields the same functional and consequently the
same stationary distribution. Our work therefore brings some support to the
conjecture that these distributions are universal. Another aim of this
work is to draw attention to the fact that the same probability distribution
also arises in other physical problems, for instance in the context of
diffusion in a random medium. 


\section{The model with a white noise potential}\label{sec:gau}

Consider the Schr\"odinger equation on the half line $x\ge0$
\be
-\frac{d^2}{dx^2}u(x)+V(x)u(x)=k^2u(x)
\:.\ee
We assume that the potential has its support on the interval $[0,L]$ and
impose Dirichlet boundary condition at the origin $u(0)=0$.
Therefore for $x>L$ scattering states of the form
$u(x)=\EXP{-ikx}+\EXP{ikx+i\theta(k)}$ represent the superposition of an
incoming plane wave and a reflected plane wave. In this case the reflection
coefficient $r(k)=\EXP{i\theta(k)}$ is of unit modulus since there is only
backward scattering.
All information on the scattering process is contained in the phase
shift $\theta(k)$. In particular the Wigner time delay is given by
\be\label{timedelay}
\tau(k)=\frac{1}{2k}\frac{d\theta(k)}{dk}
\:.\ee
Our aim is to find the probability distribution of $\tau(k)$ in the case
where the potential is a Gaussian white noise such that
\bea\label{whitenoise}
\mean{V(x)}&=&0 \nonumber\\
\mean{V(x)\,V(y)}&=&\sigg\,\delta(x-y)
\:.\eea
A widely used method to deal with such problems is the invariant embedding
approach \cite{rammal,heinrichs}. Here we will use a more straightforward
derivation following \cite{sulem}.

The Ricatti variable $z(x)=\frac{d}{dx}\ln|u(x)|$ satisfies the first
order nonlinear differential equation
\be\label{ricatti1}
\frac{dz(x)}{dx}=V(x)-k^2-z(x)^2
\:.\ee
In the region where $V(x)$ vanishes one has
\be
z(x)=\frac{-ik+ik\,\EXP{2ikx+i\theta(k)}}{1+\EXP{2ikx+i\theta(k)}}
\:.\ee
Inverting this relation we may express the reflection coefficient in terms of
the Ricatti variable
\be
\EXP{2ikx+i\theta(k)}=\frac{ik+z(x)}{ik-z(x)}
\:.\ee
Although this relation only holds in the region $x>L$, we may take it as a
definition of the variable $\psi(x)$
\be
\EXP{i\psi(x)}=\frac{ik+z(x)}{ik-z(x)}
\ee
in the region $0\le x\le L$.
By using equation (\ref{ricatti1}) one gets the differential equation
satisfied by $\psi(x)$
\be\label{psieq1}
\frac{d\psi(x)}{dx}=2k-\frac{1}{k}\Big(1+\cos\psi(x)\Big)V(x)
\:.\ee
Since the Ricatti variable is continuous even if the potential has a
discontinuity this implies that the phase shift $\theta(k)$ is given by
$\psi(L,k)-2kL$. The reflection condition at the origin gives the boundary
condition $\psi(0)=\pi$.

Let us introduce the derivative of $\psi$ with respect to $k$: 
$Z(x,k)\equiv\frac{d}{dk}\psi(x,k)$ in terms of which the time delay reads
\be\label{tauZ}
\tau(k)=\frac{Z(L,k)-2L}{2k}
\:.\ee
The differential equation for $Z$ now reads 
\be\label{Zeq1}
\frac{dZ(x)}{dx}=2+
\left[ \frac{1}{k^2} \Big(1+\cos\psi(x)\Big) + \frac{1}{k} Z(x) \sin\psi(x) 
\right] V(x)
\:.\ee
From the boundary condition at $x=0$ we may set $Z(0)=0$.

If $V(x)$ is a Gaussian white noise equations (\ref{psieq1},\ref{Zeq1}) 
are coupled stochastic differential equations (in the Stratonovich sense) 
which give the phase and the time delay.

One then may pass from these two stochastic differential equations to a
Fokker-Planck equation for the probability density of $\psi$ and $Z$.
In the high energy limit $k\gg \sigg^{1/3}$ one can show that 
the variable $\psi$ is a rapid variable uniformly distributed on the 
interval $[0,2\pi]$. 
Moreover, since the rapid variable $\psi$ and the slow variable 
$Z$ decorrelate in this limit, then one may average over the rapid variable 
and eventually get the following Fokker-Planck equation for $Z$ (a more 
detailed description of the procedure will be given in section~\ref{sec:susy})
\be\label{FPZ1}
\drond{P(Z;x)}{x}=\drond{}{Z}\left[
  \Big(\frac{\sigg}{4k^2} Z -2\Big) P(Z;x) 
\right] +
\frac{\sigg}{4k^2}\drond{}{Z} Z \drond{}{Z} Z \,P(Z;x)
\:.\ee
Up to an inessential term\footnote{
we do not write this term because it is
of the same order as terms we neglected in the approximation of the
decorrelation of $\psi$ and $Z$.
}, this equation coincides with the one derived in \cite{kumar} and 
\cite{heinrichs}.
It belongs to a more general class of Fokker-Planck
equations that appear in the study of  exponential functionals of
Brownian motion with drift.


\section{A representation of $\tau$ as an exponential functional of
the Brownian motion}

Equation (\ref{FPZ1}) describes a special case of a
class of stochastic processes which have
been studied extensively in the mathematical \cite{yor3} as well as in 
the physical literature. 
It can be cast into the equation studied by Schenzle et al. in the context
of multiplicative stochastic processes \cite{schenzle}. More recently it was
shown to arise in the context of diffusion in a random medium
\cite{BCGD,CMhyp}.
The distribution of the flux of particles \cite{oshanin,comtet} 
in a disordered sample of finite length or the waiting time distribution
may be obtained by solving a generalization of equation
(\ref{FPZ1}). From our previous work \cite{comtet,monthus} one may write 
the general solution of (\ref{FPZ1}) in the form
\be\label{Zdistr}
P(Z;L)=\frac{\lambda}{Z^2}\EXP{-\frac{\lambda}{Z}}
+\frac{2}{\pi Z}\EXP{-\frac{\lambda}{2 Z}}
\int_0^\infty ds\,\EXP{-\frac{L}{2\lambda}(1+s^2)}
\frac{s}{1+s^2}\sinh{\frac{\pi s}{2}}\ \ 
W_{1,\frac{is}{2}}\left(\frac{\lambda}{Z}\right)
\ee
where $W_{\mu,\nu}(z)$ is a Whittaker's function and
$\lambda=\frac{8k^2}{\sigg}$ is the localization length at high energy
($k\gg\sigg^{1/3}$).
In the limit $L\to\infty$ the first term gives the limiting distribution
which, as announced, coincides with the one channel distribution obtained in
\cite{fyodorov}. It is also in agreement with the result of Jayannavar et al.
\cite{kumar} in the high energy limit.

The general solution given above allows to study finite size effects.
The first correction to the stationary distribution for $Z$ may be
calculated from (\ref{Zdistr}) 
\be
P(Z;L)\APPROX{L\gg\lambda}
\frac{\lambda}{Z^2}\EXP{-\frac{\lambda}{Z}} +
\sqrt{\frac{\pi}{2}}\left(\frac{\lambda}{L}\right)^{3/2}
\EXP{-\frac{L}{2\lambda}}\ W_{1,0}\left(\frac{\lambda}{Z}\right)
\frac{\EXP{-\frac{\lambda}{2Z}}}{Z}
\ee

One can also compute all the moments of the variable $Z$ (or $\tau$)
\cite{comtet,monthus}. For $n$
greater than one they all diverge exponentially as a function of the length
of the sample. 
\bea
\mean{Z(L)}&=&2L \\
\mean{Z(L)^n}&\simeq&\frac{(n-2)!}{(2n-2)!}\lambda^n
\EXP{2n(n-1)\frac{L}{\lambda}}
\eea
It is interesting to remark that the limiting distribution
is not of the log-normal form as could have been expected \cite{heinrichs}
from a simple resummation of the most divergent part of the moments.

Another by-product of our earlier work is to provide a representation of the
process to which the random variable $\tau$ obeys. The corresponding
stochastic differential equation associated with (\ref{FPZ1}) may be written
in the following form
\be
\frac{dZ(x)}{dx}= 2-\frac{\sigg}{4k^2} Z(x) +  \frac{1}{\sqrt{2}k}Z(x) V(x)
\ee
where $V(x)$ is the white noise (\ref{whitenoise}).
By integration one obtains $Z$ and from (\ref{tauZ}) the time delay
\be\label{timedelay1}
\tau(k)=\frac{1}{k}\int_0^L dx \left(
  \EXP{\int_x^L dx'\,\left(\frac{V(x')}{\sqrt2\,k}-\frac{\sigg}{4k^2} \right)}
  -1
\right)
\:.\ee
A more satisfactory procedure  to derive this formula is to start from the 
stochastic differential equations (\ref{psieq1},\ref{Zeq1}) and construct 
$\tau$ without using a Fokker-Planck equation. This approach was 
used by Faris and Tsay \cite{faris}.

When writing this expression for $\tau(k)$ it should be kept in mind that
although (\ref{timedelay1}) is not 
true for every realization of $V(x)$, it nevertheless captures all the
statistical properties of the process.
In particular it allows the study of the various limiting cases.
Since the potential $V(x)$ is a white noise, the first term in the
exponential, which is a Brownian motion,
is typically of order $\sqrt{\sigg L/k^2}$. The argument of the
exponential is roughly a function of the ratio of the localization length
$\lambda(k)=8k^2/\sigg$ and the size of the system $L$.
In particular if $\lambda\ll L$ the time delay is approximatively equal to
$-L/k$ which is twice the length of the sample divided by the speed of the
particle. This value for the time delay corresponds to a reflection at the
sample edge. This agrees with the fact that if $\lambda\ll L$, one expects
that the particle will only enter the sample for a negligeable length
compared to $L$.
In the other regime $\lambda\gg L$ expression (\ref{timedelay1}) shows
that the time delay is roughly zero. This is consistent with the fact that 
in this regime the sample is almost transparent to the particle.

Moreover if one considers the regime where $\lambda\ll L$, the stationary
distribution for $Z(L)$ gives a typical value $\typ{Z}\simeq 2\lambda$
corresponding to a typical time delay $\typ{\tau}\simeq-\frac{L-\lambda}{k}$.
Such a value for the time delay means that the particle enters in the
disordered region for a typical length $\lambda$.

One may wonder to what extent the representation of $Z$ as an exponential
functional of the potential depends on the specific form of the noise that
we have considered. Although we are not able to give a definite answer to this
question we nevertheless notice that for any realization of $V(x)$
one may integrate equation (\ref{Zeq1}) and get
\be
Z(L)=2\int_0^L dx\,
\left(
  1+ \frac{V(x)}{2k^2} (1+\cos\psi(x))
\right)
\EXP{\frac{1}{k}\int_x^L dx'\,V(x')\sin\psi(x')}
\ee
hence
\be
\tau(k)\APPROX{k\gg\sigg^{1/3}}
\frac{1}{k}\int_0^L dx\,\left(
  \EXP{\frac{1}{k}\int_x^L dx'\,V(x')\sin\psi(x')} -1 \right)
\:.\ee
Up to a constant drift, at high energy, this is 
essentially of the form given in 
(\ref{timedelay1}). The distribution and the asymptotic behaviour of
exponential functionals of more general processes like Levy processes has
been obtained in the mathematical literature \cite{yor1,yor2}. There it is 
shown that a Poisson process  gives a limiting distribution whose 
tail still decays algebraically.


\section{The supersymmetric model}\label{sec:susy}

\subsection{The model}

In this section we consider a different model for which one can also compute
exactly the time delay distribution in the weak disorder limit. 
We show below that
$\tau(k)$ is distributed with the same law as in the previous case. 

We consider the one-dimensional Schr\"odinger Hamiltonian
\be\label{susyham}
H=-\frac{d^2}{dx^2}+\phi^2(x)+\phi'(x)
\:.\ee
As explained at length in \cite{BCGD,CDM,junker,cooper}
this model arises in diverse areas of
quantum mechanics ranging from the study of solitons in polymers to the
study of classical diffusion in a random medium.
Recently it was used in the context of one-dimensional spin chains
\cite{melin} and isospectral periodic potentials \cite{dunne}.

In the following we consider the case where $\phi(x)$ is a Gaussian white
noise with the moments
\bea
\mean{\phi(x)}&=&0\nonumber\\
\mean{\phi(x)\,\phi(y)}&=&\sigma\,\delta(x-y)
\:.\eea
The supersymmetric  Hamiltonian (\ref{susyham}) 
can be rewritten in the factorized form
\be\label{susyham2}
H=Q^\dagger Q
\ee
where
\bea
Q&=&-\frac{d}{dx}+\phi(x)\\
Q^\dagger&=&\frac{d}{dx}+\phi(x)
\:.\eea

The density of states and the localization length for this model were first 
obtained in \cite{ovchi} and then rediscovered independently in 
\cite{BCGD2}.
In contrast with the previous model,
in the  high energy limit ($k\gg\sigma$) the localization length
reaches the constant value $\lambda(k)=\frac{2}{\sigma}$. 

The two other length scales of the problem are the size of the
disorder region $L$ and the de Broglie wavelength $k^{-1}$. 
In the following  we choose to work in a high energy limit, i.e. 
when $k\gg\sigma$ and $k\gg L^{-1}$.

As in the previous section, the potential is non zero on the interval
$[0,L]$ and there is a reflection condition at the origin.

\subsection{Stochastic differential equations for the phase and the time
delay}

Using the factorization of the supersymmetric Hamiltonian, one may decouple
the stationary Schr\"odinger equation
$\left(-\frac{d^2}{dx^2}+\phi^2(x)+\phi'(x)\right)u(x)=k^2\,u(x)$
into two first order equations:
\bea\label{susyham3}
\frac{du(x)}{dx}&=&\phi(x)u(x)-kv(x) \\
\frac{dv(x)}{dx}&=&-\phi(x)v(x)+ku(x) 
\:.\eea
If $\phi(x)$ has a discontinuity the two functions $u(x)$ and $v(x)$ are 
continuous.
This suggests the introduction of the Ricatti variable 
$\zeta(x)=\frac{v(x)}{u(x)}$. 
It is straightforward to see that $\zeta(x)$ obeys the following first 
order non linear differential equation:
\be\label{ricatti}
\frac{d\zeta(x)}{dx}=k-2\phi(x)\zeta(x)+k\,\zeta(x)^2
\:.\ee
In the region where $\phi(x)$ is vanishing ($x>L$), the stationary scattering 
states can be expressed as 
\be\label{phasedef}
u(x)=\EXP{-ikx}+\EXP{ikx+i\theta(k)}
\ee
where $\theta(k)$ is the phase shift.

This leads to the change of variable:
\be
\zeta(x)=i\frac{1-\EXP{2ikx+i\alpha(x)}}{1+\EXP{2ikx+i\alpha(x)}}
\ee
with the phase shift $\theta(k)$ given by $\alpha(L,k)$.
Instead of $\alpha(x)$ it is in fact more convenient to
introduce the variable $\psi(x)=\alpha(x)+2kx$ in terms of which the 
Ricatti variable is $\zeta(x)=\tan(\psi(x)/2)$. 
Equation (\ref{ricatti}) then gives
\be\label{psieq}
\frac{d\psi(x)}{dx}= 2k-2\phi(x)\sin\psi(x)
\:.\ee
This gives the evolution of the phase and consequently the distribution of
the nodes of the wave function $u(x)$ from which one can get the density of
states (compare with equation (2.5) of \cite{CDM}).

As in the previous section one may introduce $Z(x,k)$, the derivative of 
$\psi(x,k)$ with respect to $k$, which satisfies
\be\label{Zeq}
\frac{dZ(x)}{dx}=2-2\phi(x)Z(x)\cos\psi(x)
\ee
again the initial conditions for the two variables are $\psi(0)=\pi$ and
$Z(0)=0$.

Integrating the coupled equations (\ref{psieq},\ref{Zeq}) between $0$ 
and $L$ gives the phase shift and the time delay.

\subsection{The phase distribution}

At this stage the formalism developed to compute the phase shift and the
time delay is quite general. 
We now consider the case where $\phi(x)$ is  white noise.
Equation (\ref{psieq}) is then a stochastic
differential equation in the Stratonovich sense from which we can write down
a Fokker-Planck equation for the probability density of $\psi$ :
\be\label{FPpsi}
\drond{P(\psi;x)}{x}=-2k\drond{}{\psi}P(\psi;x)+2\sigma
\drond{}{\psi}\sin\psi\drond{}{\psi}\sin\psi P(\psi;x)
\:.\ee
The stationary distribution for $\psi$ is given by:
\be\label{SDpsi}
P_{s}(\psi)={\cal N}\,\frac{\EXP{-\frac{k}{\sigma}\cotg\psi}}{\sin\psi}
\int_{\psi}^{\pi}d\psi'\frac{\EXP{\frac{k}{\sigma}\cotg\psi'}}{\sin\psi'}
\ee
where ${\cal N}$ is a normalization constant related to the integrated density
of states per unit length $N(E)=2kP_s(\pi)=2\sigma{\cal N}$ (see equation 
(2.18) of \cite{CDM} and reference therein).
One can extract from 
(\ref{SDpsi}) the high energy expansion of the stationary distribution:
\be\label{SDpsiexp}
P_{s}(\psi)={\cal N}\,\frac{\sigma}{k}\left\{
  1+\frac{\sigma}{2k}\sin2\psi+
  \frac{\sigma^2}{k^2}
  \left(\frac{1}{2} \sin^2 2\psi  - \sin^4\psi
  \right) + O\left(\frac{\sigma^3}{k^3}\right)
\right\}
\:.\ee
This shows that in the high energy limit the phase is a uniformly distributed
variable on $[0,2\pi]$. This could have been guessed directly from the 
differential equation for $\psi$ since in equation  (\ref{psieq}) one
expects that for $k\gg\sigma$ the first term will dominate.

\subsection{The time delay}

Since $\psi$ is a rapidly varying variable, one expects that the 
two coupled differential equations will reduce to only
one, after averaging over the rapid variable. For this purpose, 
the next natural assumption  is the decorrelation of
the two variables $\psi$ and $Z$ in the high energy limit. We first
integrate perturbatively equations (\ref{psieq},\ref{Zeq}) to show
that this is indeed the case:
\be
\psi(x)=\pi+2kx+2\int_0^x dx'\,\phi(x')\sin2kx'+\cdots
\ee
Since the integral is of the order $\sqrt{\sigma x}$, this expansion is
valid if $\sqrt{\sigma x}\ll 1$. The computation of the autocorrelation function 
$\mean{\psi(x)\psi(y)}-\mean{\psi(x)}\mean{\psi(y)}$
then shows that the variable $\psi$ behaves in this limit as a Brownian
motion with a drift:
\be
\psi(x)\simeq\pi+2kx+\sqrt{2}\int_0^x dx'\,\phi(x')
\:.\ee
One can perform the same approximation for $Z$ and
compute the correlation function with the help of the two expansions 
\be\label{corr}
\mean{\psi(x)Z(x)}-\mean{\psi(x)}\mean{Z(x)}\simeq
\frac{1}{k}\left(
  \frac{\sigma}{4k}\sin4kx - \sigma x\cos4kx
\right)
\:.\ee
From which we deduce that the correlations between $\psi$ and $Z$ will 
vanish in the high energy limit.

From the two stochastic differential equations (\ref{psieq},\ref{Zeq})
we may write the Fokker-Planck equation for the joint probability
density $P(\psi,Z;x)$:
\bea\label{FPpsiZ}
\drond P x = -2k \drond P \psi -2\drond P Z + 2\sigma \left\{
  \drond{}{\psi}\sin\psi\drond{}{\psi}\sin\psi \,P
  + \drond{}{\psi}\sin\psi \drond{}{Z}Z\cos\psi \,P
\right.\nonumber\\
\left.
  + \drond{}{Z}Z\cos\psi \drond{}{\psi}\sin\psi \,P
  + \drond{}{Z}Z\cos\psi \drond{}{Z}Z\cos\psi \,P
\right\}  
\:.\eea
Using the fact that $\psi$ is uniformly distributed and decorrelated from
$Z$ in the high energy limit, we may average over $\psi$ since the joint
probability density factorises as $P(\psi,Z;x)\simeq\frac{1}{2\pi}P(Z;x)$.
This leads to the following Fokker-Planck equation for $P(Z;x)$:
\be\label{FPZ}
\drond{P(Z;x)}{x}=\drond{}{Z}\left[(\sigma Z -2) P(Z;x)\right] +
\sigma\drond{}{Z} Z \drond{}{Z} Z \,P(Z;x)
\ee
which is the same as equation (\ref{FPZ1}) provided one replaces
$\frac{8k^2}{\sigg}$ by $\frac{2}{\sigma}$. In particular this implies that
the distribution for the variable $Z$ is still given by equation
(\ref{Zdistr}) provided $\lambda$ is the localization length for the
supersymmetric model. 

One may thus give a representation for the time delay, valid for both models:
\be
\tau(k)=\frac{1}{k}\int_0^L dx \left(
  \EXP{ 2 \int_x^L dx'\,(\frac{\eta(x')}{\sqrt\lambda}-\frac{1}{\lambda} )} -1
\right)
\ee
where $\eta(x)$ is the white noise of variance $1$ entering in the potential
(equal to $\frac{1}{\sqrt\sigg}V(x)$ in the first case and
$\frac{1}{\sqrt\sigma}\phi(x)$ in the second) and $\lambda$ is the
localization length.


\section{Conclusion}

We have shown in this paper that, for two models of random 
potential, the time delay exhibits the same distribution. These
statistical properties are in agreement with other approaches, underlying the
universality of such properties. 

We have also demonstrated 
that in both cases the time delay may be expressed as an
exponential functional of the white noise which enters in the potential.

It would be interesting to extend this approach to the multichannel case for
which the stationary distribution has been recently obtained
\cite{beenakker}. Another point that deserves attention is the fact that the
resulting distribution for the time delay is exactly of the same form as 
the waiting time distribution that occurs in the context of classical
diffusion in a random potential \cite{BCGD}. The supersymmetric model may 
help to explore this relation.

\section*{Acknowledgments}

We thank Eug\`ene Bogomolny, Pierre Le Doussal, Rodolfo Jalabert, 
C\'ecile Monthus and Marc Yor for interesting discussions. 
We thank Niall Wheelan for reading the manuscript.


\end{document}